\documentclass[onecolumn,aps,showpacs,showkeys,preprint]{revtex4}
\usepackage{tipa}
\usepackage{bbding}
\usepackage{txfonts}
\usepackage{amssymb}
\usepackage{graphicx}

\begin{document}

\title{Patterning graphene nanostripes in substrate-supported functionalized graphene: A promising route to integrated, robust, and superior transistors}

\author{Liang-feng HUANG}
\author{Zhi ZENG}\email{zzeng@theory.issp.ac.cn}\affiliation{Key
Laboratory of Materials Physics, Institute of Solid State Physics,
Chinese Academy of Sciences, Hefei 230031, People's Republic of
China}

\begin{abstract}
It is promising to apply quantum-mechanically confined graphene
systems in field-effect transistors. High stability, superior
performance, and large-scale integration are the main challenges
facing the practical application of graphene transistors. Our
understandings of the adatom-graphene interaction combined with
recent progress in the nanofabrication technology indicate that very
stable and high-quality graphene nanostripes could be integrated in
substrate-supported functionalized (hydrogenated or fluorinated)
graphene using electron-beam lithography. We also propose that
parallelizing a couple of graphene nanostripes in a transistor
should be preferred for practical application, which is also very
useful for transistors based on graphene nanoribbon.
\end{abstract}

\preprint{Perspective}

\keywords{Graphene nanostripe, functionalized graphene, field-effect
transistor}
\pacs{61.48.Gh, 68.65.-k, 85.30.Tv}

\maketitle

\par Since the first fabrication and measurement of graphene in 2004\cite{novoselov306}, it has become
promising in the fields of materials, chemistry, physics, and
biology, due to its superior mechanical, electronic and optical
properties, low mass density, controllable synthesis, biological
compatibility, and so
on\cite{geim6,wei22,rozhkov503,singh56,molitor23,stampfer63,sheng_2011,bao19,liu474,bao5}.
Graphene transistor is an important application of graphene in these
fields, which recently has drawn tremendous interest from both the
scientific and industrial communities. The bulk graphene is a
semi-metal, and a band gap should be created to achieve high on/off
(signal/noise) ratio in graphene transistors\cite{schwierz5}.

\subsection{High stability, atomically sharp edges, substrate
decoupling, and large-scale integration are required for the
practical application of graphene nanostructures in field-effect
transistors}

\par Quantum-mechanically confining the carriers in graphene nanoribbon (GNR) is an effective and
controllable way for the band-gap
engineering\cite{chen40,han98,wang100,li319,jiao321,tao7}. GNRs
currently have been fabricated through masked plasma
etching\cite{chen40,han98}, chemical derivation\cite{wang100,li319},
unzipping carbon nanotubes\cite{jiao321,tao7}, nanoparticle
etching\cite{campos9}, surface-assisted self-assembly\cite{cai466},
templated self-organization\cite{sprinkle5}, and nanoimprint
lithography\cite{liang10}. The band gap of GNR increases inversely
with the ribbon width, and a sub-10-nm width is required for a
satisfactory on/off ratio in a GNR-based field-effect transistor
(GNR-FET) under conventional
conditions\cite{han98,li319,liang10,tao7}. Presently, only the
masked-plasma etching method has successfully integrated sub-10-nm
GNRs on semiconductor substrate\cite{yang22}. However, the
electronic performance of narrow GNRs is still significantly
affected by the edge disorder (caused by the high-energy plasma
etching), chemical bonding with unexpected adsorbates, and many
detrimental effects from the substrate (additional carrier
scattering, intrinsic charge doping, parasitic capacity, and
unpredictable disorder-induced
gap)\cite{wang100,koskinen80,wang324,gallagher81,wang4,xu11}.
Probably due to some edge disorder\cite{bai5}, the theoretically
predicted spin-polarized transport in zigzag-GNR\cite{kim3} has not
been experimentally observed. Thus, further improvement in the
fabrication of GNR is still required to overcome these obstacles for
its practical usage. In addition, large-scale integration is also
required for the application of GNR in semiconductor industry.

\par
\begin{figure}[p]
\scalebox{1.1}[1.1]{\includegraphics{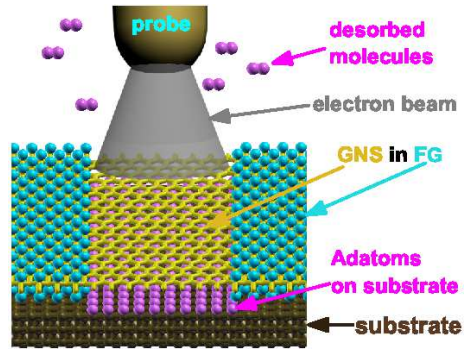}}
\caption{\label{STM_sculpture} The schematic drawing of the
sculpturing of a GNS in a substrate-supported FG sheet by
electron-beam lithography. The electron beam is generated using a
biased scanning probe.}
\end{figure}

\begin{figure}[h]
\scalebox{0.6}[0.6]{\includegraphics{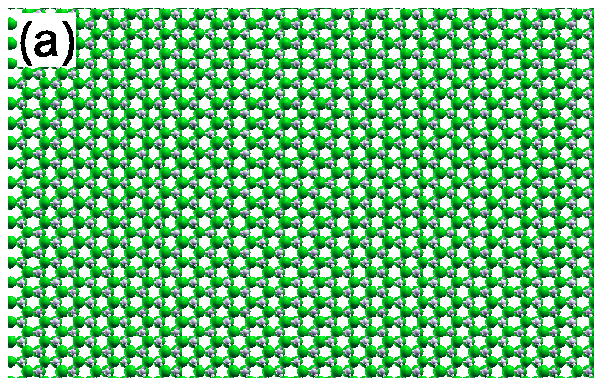}}
\scalebox{0.6}[0.6]{\includegraphics{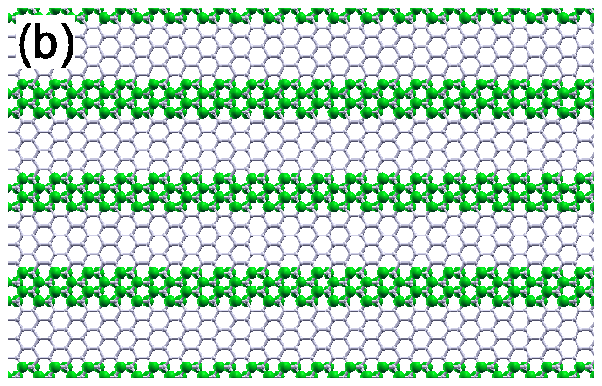}}
\scalebox{0.6}[0.6]{\includegraphics{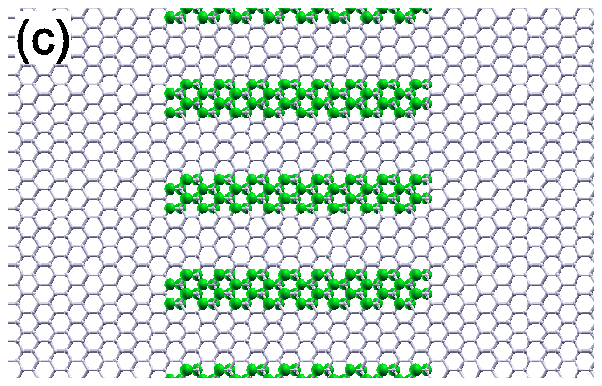}}
\scalebox{0.6}[0.6]{\includegraphics{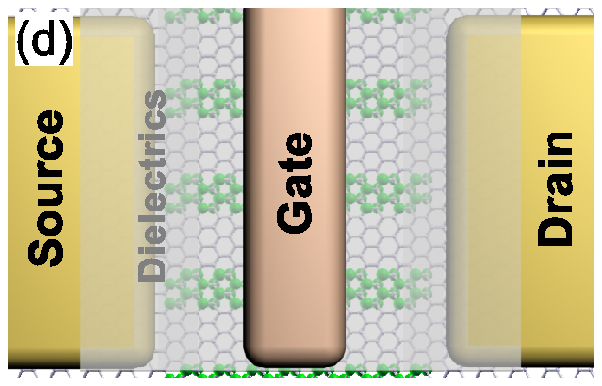}}
\scalebox{0.6}[0.6]{\includegraphics{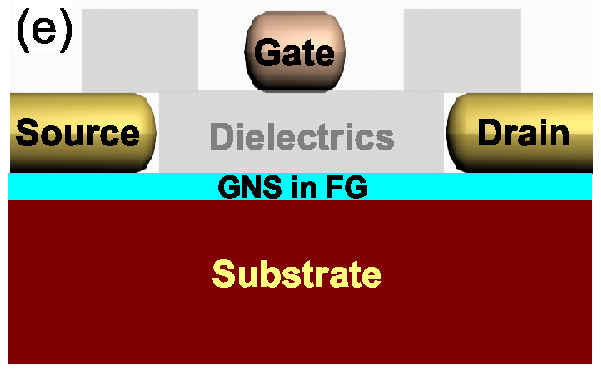}}
\caption{\label{GNS_FET} The schematic drawings of (a--d) the
fabrication steps of a GNS-FET, which consists of a couple of
parallelized GNSs, and (e) the side view of a GNS-FET. The width of
the GNSs should be sub-10 nm.}
\end{figure}

 As a counterpart of GNR, the quasi-one-dimensional graphene nanostripe (GNS) sculptured
in functionalized (hydrogenated or fluorinated) graphene (FG) has
the similar electronic structure as that of GNR, and is regarded as
another promising candidate for
FET\cite{singh9,singh4,munoz19,ribas4,huang_GNS}. To obtain superior
electronic performance, the edges of GNS, namely the graphene/FG
interfaces, should be sharp enough down to the atomic scale, and the
fabricated GNS should be robust (without structural changing for a
very long time) under conventional conditions. Furthermore, to be an
alternative or even a substitute of present silicon-based FETs,
large-scale integration of GNS-FET should be readily realized.
Scanning-probe lithography is a powerful and miniature electron-beam
lithography, which can fabricate nanostructures on surfaces with
dimensions even down to several nanometers\cite{tseng23}. This
method could be used to sculpture sub-10-nm GNSs in FG. The
sculpturing process is designed as shown in Fig.
\ref{STM_sculpture}, in which a GNS is sculptured out in a
substrate-supported FG sheet by using a biased scanning probe. The
C--X bonds (X is H or F) at the graphene/FG interfaces are broken by
the electron-beam excitation. The molecules desorbed upward will
escape into vacuum with high escaping velocity, while those downward
tend to dissociate again and be adsorbed onto the semiconductor
substrate\cite{huang_GNS}. Those adatoms on the substrate prevent
the GNS from covalently bonding with the
substrate\cite{riedl103,robinson11,wong_acsnano}, which can
guarantee the superior electronic performance of the fabricated GNS.
Through systematically simulating the kinetics of the adatoms at
various graphene/FG interfaces\cite{huang_GNS}, we have shown that
it is probable to sculpture GNSs with sharp and highly-stable edges
in FG. This is because of the considerable increase of the
reactivity of the C atoms to H (or F) adatoms with changing the
adsorption site from $sp^2$-hybridized GNS to
$sp^3$-FG\cite{huang_GNS,ao97,lin78}. Furthermore, scanning-probe
lithography should be capable of integrating GNSs in
FG\cite{tseng23}, which is schematically shown in Fig.
\ref{GNS_FET}(a) and (b). Neighboring GNSs can be easily decoupled
by just couples of adatom lines\cite{lee97}, which indicates that
the GNSs could be sculptured very close to but without influencing
each other. Thus, high-density integration of GNSs can be realized
in FG. In addition, the edges of the experimentally fabricated GNSs
in hydrogenated\cite{sessi9} and fluorinated graphene\cite{withers}
are still not atomically sharp, which is probably due to the fact
that the adatom coverage is not high enough in these two
experiments\cite{huang_GNS}. Thus, to obtain sharp-edge GNSs, it is
critical to have FG sheets with high enough adatom coverage.

\subsection{Parallelizing a couple of GNSs in a FET can make the
integrated GNS-FETs perform uniformly and be immune to small
structural and environmental perturbations, which also could be
generalized to GNR-FET}

\par
\begin{figure}[h]
\scalebox{0.6}[0.6]{\includegraphics{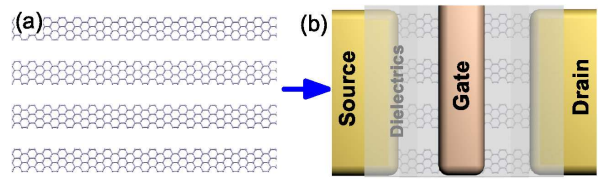}}
\caption{\label{GNR_FET} Parallelizing a couple of GNRs in a GNR-FET
could also make all the FETs in an integrated circuit perform more
uniformly than those each consists only of one GNR. The width of the
GNRs should be sub-10 nm.}
\end{figure}

 After sculpturing a GNS in a substrate-supported FG sheet,
a GNS-FET can be achieved by depositing the source and drain
electrodes, dielectrics, and gate electrode, subsequently. However,
one important aspect needs a special concern, which is that the
electronic correlation in this atomically thick quasi-one
dimensional GNS should be quite strong, especially for very narrow
GNS. Thus, any small variation in the structure and/or environment
will possibly result in an obvious discrepancy in the transport
property between different GNSs. This issue may become harmful in
large-scale integrated circuits if the unit device (GNS-FET) only
consists of one GNS, because it is impossible that the structure and
environment of different GNSs in the circuit be strictly all the
same. This will introduce additional noise and limit the circuit
performance. However, when parallelizing a couple of GNSs in a FET,
the variation in the structure, environment, and then the transport
property can be averaged out, and the noise will be lowered. The
preparation steps for a GNS-FET consisting of a couple of GNSs are
schematically shown in Fig. \ref{GNS_FET}(a--e). This
parallelization of several transport channels in a FET can be
generalized to GNR-FET, too, which is schematically shown in Fig.
\ref{GNR_FET}. We could foresee that this parallelization approach
can make the integrated FETs perform more uniformly and be immune to
small structural and environmental perturbations, and should be
preferred for high-quality graphene-based circuits.

\par In summary, GNS sculptured in FG
is a promising nanostructure for graphene transistor. Scanning-probe
lithography can be used to obtain integrated, robust, and superior
GNSs in substrate-supported FG. It is possible for GNS to overcome
some obstacles currently facing the practical application of GNR.
Parallelizing a couple of GNSs in a GNS-FET could make all the FETs
in an integrated circuit perform uniformly and be immune to the
structural and environmental perturbations, which also can be
generalized to GNR-FET.

\subsection*{Acknowledgement} This work is supported by the National
Science Foundation of China under Grant No. 11174284, National Basic
Research Program of China (973 Program) under Grant No.
2012CB933702, Knowledge Innovation Program of Chinese Academy of
Sciences, and Director Grants of CASHIPS.

\bibliography{lfhuang_bib}

\begin{thebibliography}{10}

\bibitem{tseng23}
T.~P.~Chen A.~A.~Tseng, A.~Notargiacomo.
\newblock {\em J. Vac. Technol. B}, 23:877, 2005.

\bibitem{geim6}
K.~S.~Novoselov A.~K.~Geim.
\newblock {\em Nat. Mater.}, 6:183, 2007.

\bibitem{singh9}
B.~I.~Yakobson A.~K.~Singh.
\newblock {\em Nano Lett.}, 9:1540, 2009.

\bibitem{singh4}
B.~I.~Yakobson A.~K.~Singh, E. S.~Penev.
\newblock {\em ACS Nano}, 4:3510, 2010.

\bibitem{rozhkov503}
Y.~P. Bliokh V. Freilikher F.~Nori A.~V.~Rozhkov, G.~Giavaras.
\newblock {\em Phys. Rep.}, 503:77, 2011.

\bibitem{riedl103}
T.~Iwasaki A. A. Zakharov U.~Starke C.~Riedl, C.~Coletti.
\newblock {\em Phys. Rev. Lett.}, 103:246804, 2009.

\bibitem{stampfer63}
J.~G\"uttinger F. Molitor C. Volk B. Terr\'es J. Dauber S. Engels S. Schnez A.
  Jacobsen S. Dr\"oscher T. Ihn K.~Ensslin C.~Stampfer, S.~Fringes.
\newblock {\em Front. Phys.}, 6:271, 2011.

\bibitem{tao7}
O.~V. Yazyev Y. C. Chen J. Feng X. Zhang R. B. Capaz J. M. Tour A. Zettl S. G.
  Louie S. G. H. Dai M. F.~Crommie C.~Tao, L.~Jiao.
\newblock {\em Nat. Phys.}, 7:616, 2011.

\bibitem{wei22}
Y.~Q.~Liu D.~C.~Wei.
\newblock {\em Adv. Mater.}, 22:3225, 2010.

\bibitem{munoz19}
M.~A. Ribas E. S. Penev B. I.~Yakobson E.~Mu\~noz, A. K.~Singh.
\newblock {\em Diamond \& Related Materials}, 19:368, 2010.

\bibitem{molitor23}
C.~Stampfer S. Dr\"oscher A. Jacobsen T. Ihn K.~Ensslin F.~Molitor,
  J.~G\"uttinger.
\newblock {\em J. Phys.: Condens. Matter}, 23:243201, 2011.

\bibitem{withers}
M.~Dubois S. Russo M.~Craciun F.~Withers, T. H.~Bointon.
\newblock {\em Nano Lett.}, 11:3912, 2011.

\bibitem{xu11}
Jr. J. Tang J. Bai E. B. Song Y. Huang X. Duan Y. Zhang K. L.~Wang G.~Xu, C.
  M.~Torres.
\newblock {\em Nano Lett.}, 11:1082, 2011.

\bibitem{wang4}
C.~Cong J. Shang T.~Yu H.~Wang, Y.~Wu.
\newblock {\em ACS Nano}, 4:7221, 2010.

\bibitem{robinson11}
M.~Labella K. A. Trumbull R. Cavelero D. W.~Snyder J.~A.~Robinson,
  M.~Hollander.
\newblock {\em Nano Lett.}, 11:3875, 2011.

\bibitem{bai5}
F.~Xiu L. Liao M. Wang A. Shailos K. L. Wang Y. Huang X.~Duan J.~Bai, R.~Cheng.
\newblock {\em Nat. Nanotechnol.}, 5:655, 2010.

\bibitem{cai466}
R.~Jaafar M. Bieri T. Braun S. Blankenburg M. Muoth A. P. Seitsonen M. Saleh X.
  Feng K. M\"ullen R. Fasel~R. J.~Cai, P.~Ruffieux.
\newblock {\em Nature}, 466:470, 2010.

\bibitem{lee97}
J.~C.~Grossman J.~H.~Lee.
\newblock {\em Appl. Phys. Lett.}, 97:133102, 2010.

\bibitem{novoselov306}
S.~V. Morozov D. Jiang Y. Zhang S. V. Dubonos I. V. Grigorieva A. A.~Firsov
  K.~S.~Novoselov, A. K.~Geim.
\newblock {\em Science}, 306:666, 2004.

\bibitem{campos9}
J.~D. Sanchez-Yamagishi J. Kong P. Jarillo-Herro L.~C.~Campos, V.
  R.~Manfrinato.
\newblock {\em Nano Lett.}, 9:2600, 2009.

\bibitem{huang_GNS}
G.~R. Zhang-L. L. Li Z.~Zeng L.~F.~Huang, X. H.~Zheng.
\newblock {\em J. Phys. Chem. C}, 115:21088, 2011.

\bibitem{jiao321}
G.~Diankov H. Wang H.~Dai L.~Jiao, X.~Wang.
\newblock {\em Nat. Nanotechnol.}, 5:321, 2010.

\bibitem{ribas4}
P.~B. Sorokin B. I.~Yakobson M.~A.~Ribas, A. K.~Singh.
\newblock {\em Nano Res.}, 4:143, 2011.

\bibitem{han98}
Y.~Zhang P.~Kim M.~Han, B.~\"Ozyilmaz.
\newblock {\em Phys. Rev. Lett.}, 98:206805, 2007.

\bibitem{liu474}
E.~Ulin-Avila B. Geng T. Zentgraf L. Ju F. Wang X.~Zhang M.~Liu, X.~Yin.
\newblock {\em Nature}, 474:64, 2011.

\bibitem{sprinkle5}
Y.~Hu-J. Hankinson M. Rubio-Roy B. Zhang X. Wu C.~Berger M.~Sprinkle, M.~Ruan.
\newblock {\em Nat. Nanotechnol.}, 5:727, 2010.

\bibitem{gallagher81}
D.~Goldhaber-Gordon P.~Gallagher, K.~Todd.
\newblock {\em Phys. Rev. B}, 81:115409, 2010.

\bibitem{koskinen80}
H.~H\"akkinen P.~Koskinen, S.~Malola.
\newblock {\em Phys. Rev. B}, 80:073401, 2009.

\bibitem{sessi9}
M.~Bode N. P.~Guisinger P.~Sessi, J. R.~Guest.
\newblock {\em Nano Lett.}, 9:4343, 2009.

\bibitem{bao5}
B.~Wang Z. Ni C. Haley-Y. X. Lim Y. Wang D. Y. Tang K. P.~Loh Q.~Bao, H.~Zhang.
\newblock {\em Nature}, 5:411, 2011.

\bibitem{bao19}
Y.~Wang Z. Ni Y. Yan-Z. X. Shen K. P. Loh D. Y.~Tang Q.~Bao, H.~Zhang.
\newblock {\em Adv. Funct. Mater.}, 19:3077, 2009.

\bibitem{yang22}
Y.~Wang Z. Shi D. Shi-H. Gao E. Wang G.~Zhang R.~Yang, L.~Zhang.
\newblock {\em Adv. Mater.}, 22:4014, 2010.

\bibitem{wong_acsnano}
Y.~Wang L. Cao D. Qi I. Santoso W. Chen A. T. S.~Wee S.~L.~Wong, H.~Huang.
\newblock {\em ACS Nano}, 5:7662, 2011.

\bibitem{schwierz5}
F.~Schwierz.
\newblock {\em Nat. Nanotechnol.}, 5:487, 2010.

\bibitem{singh56}
L.~Zhai S. Das S. I. Khondaker S.~Seal V.~Singh, D.~Joung.
\newblock {\em Prog. Mater. Sci.}, 56:1178, 2011.

\bibitem{sheng_2011}
A.~D. G\"ucl\"u M. Zielinski P. Potasz E. S. Kadantsev O. Voznyy P.~Hawrylak
  W.~D.~Sheng, M.~Korkusinski.
\newblock {\em Front. Phys.}, 2011.

\bibitem{kim3}
A.~K. S.~Kim W.~Y.~Kim.
\newblock {\em Nat. Nanotechnol.}, 3:408, 2008.

\bibitem{li319}
L.~Zhang S. Lee H.~Dai X.~Li, X.~Wang.
\newblock {\em Science}, 319:1229, 2008.

\bibitem{liang10}
S.~Wu A. Ismach D. L. Olynick S. Cabrini J.~Bokor X.~Liang, Y. S.~Jung.
\newblock {\em Nano Lett.}, 10:2454, 2010.

\bibitem{wang324}
L.~Zhang Y. Yoon P. K. Weber H. Wang J. Guo H.~Dai X.~Wang, X.~Li.
\newblock {\em Science}, 324:768, 2009.

\bibitem{wang100}
X.~Li H. Wang J. Guo H.~Dai X.~Wang, Y.~Ouyang.
\newblock {\em Phys. Rev. Lett.}, 100:206803, 2008.

\bibitem{lin78}
B.~I.~Yakobson Y.~Lin, F.~Ding.
\newblock {\em Phys. Rev. B}, 78:041402, 2008.

\bibitem{chen40}
M.~J. Rooks P.~Avouris Z.~Chen, Y. M.~Lin.
\newblock {\em Physica E}, 40:228, 2007.

\bibitem{ao97}
F.~M. Peeters S.~Li Z.~M.~Ao, A. D. Nern\'andez-Nieves.
\newblock {\em Appl. Phys. Lett.}, 97:233109, 2010.

\end{thebibliography}
\bibliographystyle{plain}


\end{document}